\documentclass[journal]{IEEEtran}
\def\BibTeX{{\rm B\kern-.05em{\sc i\kern-.025em b}\kern-.08em T\kern-.1667em\lower.7ex\hbox{E}\kern-.125emX}}

\usepackage{times}
\usepackage{gensymb}
\usepackage{tabularx}
\usepackage{lipsum}
\usepackage{array}
\usepackage{booktabs}
\usepackage{listings}
 
\usepackage{enumerate}

\usepackage{multirow}

\ifCLASSINFOpdf
 \usepackage[pdftex]{graphicx}
 \else
 \usepackage[dvips]{graphicx}

 \usepackage{amssymb,amsmath}
 
\fi

\usepackage[caption]{subfig}
\usepackage{cite}

\lstdefinestyle{lststyle}{
 captionpos=b, 
 tabsize=2,
 basicstyle=\linespread{0.9}\footnotesize\ttfamily,
}
 
\begin{document}

\title{Public Key Reinforced Blockchain Platform for Fog-IoT Network System Administration}

\author{Marc~Jayson~Baucas,~\IEEEmembership{Student Member,~IEEE}, Petros~Spachos,~\IEEEmembership{Senior Member,~IEEE}, and Konstantinos~N.~Plataniotis,~\IEEEmembership{Fellow,~IEEE} 
                 
\thanks{This work was supported in part by the Natural Sciences and Engineering Research Council (NSERC) of Canada.
\par M. Baucas and P. Spachos are with the School of Engineering, University of Guelph, Guelph, ON, N1G2W1, Canada.
(e-mail: baucas@uoguelph.ca; petros@uoguelph.ca).
\par K. Plataniotis is with the Department of Electrical and Computer Engineering, University of Toronto, Toronto, ON, M5S3G4, Canada. (e-mail: kostas@ece.utoronto.ca)}}

\maketitle
\begin{abstract}
The number of embedded devices that connect to a wireless network has been growing for the past decade. This interaction creates a network of Internet of Things (IoT) devices where data travel continuously. With the increase of devices and the need for the network to extend via fog computing, we have fog-based IoT networks. However, with more endpoints introduced to it, the network becomes open to malicious attackers. This work attempts to protect fog-based IoT networks by creating a platform that secures the endpoints through public-key encryption. The servers are allowed to mask the data packets shared within the network. To be able to track all of the encryption processes, we incorporated the use of permissioned blockchains. This technology completes the security layer by providing an immutable and automated data structure to function as a hyper ledger for the network. Each data transaction incorporates a handshake mechanism with the use of a public key pair. This design guarantees that only devices that have proper access through the keys can use the network. Hence, management is made convenient and secure. The implementation of this platform is through a wireless server-client architecture to simulate the data transactions between devices. The conducted qualitative tests provide an in-depth feasibility investigation on the network's levels of security. The results show the validity of the design as a means of fortifying the network against endpoint attacks. 
\end{abstract}


\IEEEpeerreviewmaketitle

\section{Introduction}
\noindent
\IEEEPARstart{A}{n} Internet of Things (IoT) network usually consists of many wirelessly connected devices with a common server~\cite{iot}. As networks grew, variations in the IoT architecture emerged. One example is called the fog-based IoT network. It is an IoT architecture that makes use of fog computing to extend the network and expand its scope~\cite{fog-survey}. The design uses fog devices to offload network tasks from the central servers locally. By doing so, there is an increase in the processing capacity of the central server, and network traffic is more manageable~\cite{iot-fog-sensing}. 

However, even though Fog-IoT networks optimize the standard architecture through decentralization and reallocation, it also introduces vulnerabilities in its security. In this case, it is the introduction of local servers~\cite{iot-fog-sensing}. Adding a new device to the structure creates potential entry points for malicious attacks. Also, local servers having similar functionalities and capabilities are just as crucial for the network domain. For instance, a medical centre that integrates Fog-IoT into its system can extend its services further into more remote areas~\cite{fog-med}. By taking advantage of the scope that fog computing can enable the IoT network, more users from farther places can access its health services remotely~\cite{fog-health-assist}. However, the further the connection between the server and user, the harder it is to maintain proper security of its communication. Also, with the strategic placing of local servers away from the central server, it adds another layer to the security that needs to be maintained~\cite{fog-app}. With everyone's personal information such as; health records and test results being stored digitally in a network, adding more vulnerable endpoints poses a threat to their privacy. By focusing on reinforcing these endpoints, a Fog-IoT network can remain optimized for network system administration while being secure from malicious attacks~\cite{fog-issue}.

This work proposes a platform that can protect these endpoints by regulating the devices that can access the shared data. As such, we present through this journal the following contributions:
\begin{itemize}
    \item An overview of blockchains in Fog-IoT, its uses and advantages towards improving the network.
    \item A wireless network platform that focuses on reducing the current security vulnerabilities in Fog-IoT network system administration. We chose to incorporate blockchain technology, which is known for its strong immutability and automation~\cite{block-accept} as the central data structure of the network. The server will use it to store and regulate data transactions within the platform. 
    \item The proposed platform uses permissioned blockchains in aiding Fog-IoT network security. Most blockchain-based cryptocurrency implementations use Bitcoin and Ethereum due to their security towards tampering~\cite{b-based}. However, our platform uses a less common variation called a permissioned blockchain. This design choice allows us to create an architecture that only recognizes devices through a trusted authentication~\cite{block-priv}. By using this variation of blockchain technology in our platform, we can create an authorization system that will only grant network data access to pre-authorized devices~\cite{block-access}. 
    \item A discussion on the different cryptographic techniques used against IoT network attacks such as; public-key encryption and one-way hashing. Also, we talked about their contributions towards securing that data within the server. 
    \item A data packet encryption structure using a combination of public key and one-way hashing cryptographical techniques to reinforce the data exchange within the network. This design will allow the blockchain to secure the network by making data only accessible to authorized devices. 
\end{itemize}  

The rest of this paper is as follows: a detailed discussion of each process used in the implementation in Section~\ref{back-inf}. The described experimental design and methodology of the architecture in Section~\ref{exp}. The security analysis and the evaluation are in Section~\ref{eval}. Finally, conclusions are in Section~\ref{con}.

\section{Background and Related Works} \label{back-inf}

\subsection{Security in Fog-IoT}
With the growing number of devices, a network needs to expand its scope. One of the resulting architectures makes use of fog computing to extend the network. By reallocating resources closer to the devices through a local server, more services become accessible~\cite{fog-enable-iot}. As a result, more users can effectively connect to the network without connection constraints. However, although it is easier to share data, it exposes the network to potential attacks through these newly introduced endpoints. Examples of these attacks are man-in-the-middle attacks and spoofing attacks~\cite{fog-issue}. A man-in-the-middle attack is when a third-party user can obtain shared data between two devices through a compromised transmission~\cite{mitm}. By attacking the communication medium, this attacker can indirectly eavesdrop on the data exchange without being detected. As for spoofing attacks, it is when an attacker disguises itself as a trusted device and gains authorized access to the network~\cite{spoofing}. If the malicious user is seen as a trusted device by the network, it gives access to its data. These attacks show the ability of attackers to take advantage of a vulnerable endpoint. Therefore, with the nature of a fog-based IoT network having a higher number of endpoints.

A standard IoT network security layer consists of four foundations that govern the vulnerabilities in a standard IoT network~\cite{signal}. These are endpoint protection, security monitoring and analysis, security configuration management, and communication and connectivity protection. By focusing on one of these foundations, we can increase the security of any network. We chose to focus on endpoint security since, in a fog-based IoT network, there is more data exposure along the edge of the network~\cite{fog-access-control}. This issue opens an avenue for malicious attacks. The platform proposed in this paper attempts to reinforce these endpoints by creating a more secure communication layer for the transmitted data within the network. This design is made possible by integrating blockchain technology into the Fog-IoT server infrastructure.

\subsection{Blockchain Technology}
A blockchain is a form of data structure that is copied and shared among multiple nodes~\cite{block-surv}. Each chain is a linked list of blocks with a structural design based on the data needed to be stored. However, this structure is unique. Blocks can only be added to the chain and cannot be taken from it once added. This design makes a blockchain an immutable data structure that contains time-stamped records that are protected cryptographically~\cite{b-access-priv}. Since a blockchain is shared, multiple distributed copies of the chain where a network of chains can cross-check each other to verify that all chains are the same. Before each built block, all participants decide based on a consensus type of voting. This voting system chooses where the data is stored. To be able to cast a vote, a transaction fee needs to be paid by each participant. This process is called ``mining", and the participants in the voting system are ``miners"~\cite{b-scale}. With these, blockchain technology appears to be a novel idea that can compete with current data-management systems. However, some traditional database technologies can still outperform them based on their usage. This paper chose to use blockchains due to their immutability.

A blockchain is an immutable data structure. Most records within its ledger are made irreversible with one-way cryptographic functions~\cite{b-tech}. Immutable means that the transaction history will be hard to change, which creates a reliable structure between nodes. This feature generates trust within the system since the shared records are immutable and only validated through a consensus. Unless there is a majority in control, tampering with any of the data will be difficult. This system makes this structure stronger than most traditional databases due to its unique security against attacks. Setting within the Fog-IoT network can lead to a loss of control over its data and users. Inconsistencies created by tampered network settings will lead to the whole network having unpredictable behaviour. To avoid this, the immutability of data within a blockchain can help maintain the network's integrity. As a result, it is harder to manipulate the data in a blockchain. All sensitive information is in a structure that is hard to modify. However, most know blockchains through their proof of work system. The device that has the most processing power gains access to the structure and its data. Our paper decided to choose another form of blockchain that compliments our intentions to overcome this reliance. There are two general types of blockchains, which are public and permissioned~\cite{b-based}.

\subsubsection{Public Blockchain}
A public blockchain is a type that does not need to have a trusted authority to manage its data. However, due to this lack of trust, the database is slower and less efficient. This issue is because of the need to coordinate with many anonymous participants and have no affiliation. As a result, there is an introduction of uncertainties upon implementing a public blockchain. Some of these uncertainties are; Who updates the software or firmware within the network? How much does it require to add a transaction to a block and add it to the chain? How long does it take for a decision to be made when appending to a chain? How credible are these participants? Public blockchains decide whether a device can participate in the voting process once it processes a given complex algorithm. This protocol sets demand for high processing power from its miners. This type of blockchain is less suitable for IoT network implementations due to the low-end nature of most IoT devices and sensors. Therefore, it is more preferred for developers to use permissioned blockchains instead~\cite{block-iot}.

\subsubsection{Permissioned Blockchain}
A permissioned blockchain uses a trusted protocol that pre-authorizes the devices that are allowed to access its data~\cite{b-scale}. This protocol filters these users through an access layer. As a result, only devices permitted by the ledger through a registered ID can gain control over the blockchain and its services. In comparison to its public variant, this structure provides a better means of managing its users. However, due to its tendency to have a more fixed number of miners, it reveals to be less dynamic. By having its trusted devices predefined, adding them will not be as easy as public blockchains. Although this limitation makes it seem more of a constrained version of the public blockchain, it is a trade-off between data security and user throughput ~\cite{b-fog}. Public blockchains can cater to more collections of devices but limit their users due to their high demand for processing power. However, permissioned blockchains are better at securing their users but limited by their capacity due to their trusted protocol. However, permissioned blockchains are better for IoT networks due to their advantages in data privacy~\cite{block-priv}. Therefore, a permissioned ledger allows for a more secure alternative to controlling data since information is only open to a select number of participants~\cite{block-surv}.

However, data is visible to anyone who has access to the network. A restriction still needs to be established toward unwanted endpoints within the server. Once an endpoint gains access to the network via the permissioned ledger, data is made accessible. To reinforce the security of the data and entities within the platform, we turn to cryptography. This integration can limit the privacy issues of transparency.

\subsection{Public Key Encryption}
Cryptography is a way of creating schemes and protocols that ensure secure communication between users~\cite{crypto}. In IoT, without cryptography, most networks are insecure. This issue means that there is no guarantee that transmissions are private and authentic. Cryptography allows a secure communication layer around wireless networks such as IoT servers. In this proposal, the platform uses a combination of; public-key, symmetric, and one-way hashing cryptographical techniques.

Public-key, or asymmetric cryptography, is based on each device having a key pair~\cite{public-key}. These keys are related to each other mathematically through a form of one-way encryption but are not identical. A key pair from asymmetric cryptography is composed of a public and private key. Private keys are generated into public keys using an encryption method chosen by the architecture. The public version is revealed to everyone in the network, while the private key is kept secure within each device. With proper encryption, it should be impossible to decrypt a public key to discover its private key. However, most key pairs are too big to be remembered by their users. As a result, they are stored securely within different data storage appropriated for every application that uses the technique. For example, IoT systems that use public key pairs to protect their data use the database within the cloud to store their keys~\cite{pkey-encrypt}. In decrypting a message, its contents are revealed with the decrypted message to verify its validity. It maintains the integrity of the message from the recipient to the sender and vice-versa. Each action to encrypt or decrypt data comes with a cost or resource for any processor. Also, a collection of keys can become very complex once a large number of each is stored. Therefore, network implementations of the design make use of different optimization techniques such as fast and parallel searching algorithms to manage the keys~\cite{fast-encrypt}. For our platform, we plan to use permissioned blockchains. With its automatability, we can handle the different cryptographical techniques integrated.

Another way of using the key pair in IoT is through digital signatures~\cite{digsig-iot}, which this platform incorporates. A user can sign their messages and contents through their private key. Digital signatures can provide authentication schemes that maintain the integrity and non-repudiability of data~\cite{digsig-iot-sec}. With secure authentication, only the owners of the signature can sign their messages. Also, contents cannot be invalidated by its sender since only they can sign their content. Once a public key verifies a signature, the recipient can check if the message is the same compared to its original. By doing so, there is integrity in its contents.

However, the exposure of public keys poses a vulnerability to the platform. We chose to integrate a second cryptographical tool to strengthen the security of the key pairs. Cryptographical tools can be in conjunction with one another~\cite{crypto}. The strengths of one method can cover the weaknesses of another. As a result, to verify that these secret keys and public-private key pairs are not visible to anyone who gains access to the device, one-way hashing is used. One-way hashing or digital watermarking is a mathematical algorithm that masks the actual values of a provided key~\cite{one-way-hash}. Storing passwords for large databases uses this tool to prevent potential breaches that could lead to data loss. This hashing function creates an irreversible variation of the password. Due to the algorithm's complexity, the generated variant guarantees a high chance of being a unique value. This high probability is due to its very high collision resistance. A collision occurs when two different generated hashed values have the same hashed result. Hashing functions can minimize collisions if their design is adequate. Also, most implementations of these hashing functions contain a cross-checking functionality to check if a password entered is equivalent to a provided hashed string~\cite{pkey-iot}. This design allows most databases to store these hashes without any worry of revealing sensitive information about their users. 

By applying the same construct in this paper, the construct protects the encrypted keys from malicious attackers. Having a diverse set of cryptographical tools can over-complicate a design. However, it can also create a more secure architecture that will provide better resistance to potential attackers. To create a manageable hashing tool design, we combined one-way hashing and public-key encryption.

\section{Design and Methodology} \label{exp}

\subsection{Architecture and Design Overview}
The platform focuses on the data transactions between the users and the local servers within a Fog-IoT network. It takes these two endpoints and reinforces the security of their data exchanges by using public-key encryption. We will store these keys within the private blockchain. Each blockchain will function as a secure hyper ledger. The connection between a user and the server is reinforced by placing the encrypted keys within the blockchain. Recording any data transactions within the network is also made possible for monitoring purposes. This platform allows the server to detect any malicious activity in the network to keep user data secure. Our design divides the architecture into two sections; the server and the client. 

\subsubsection{Server}
\begin{figure}[t!]
\centering
\includegraphics[width=0.7\columnwidth]{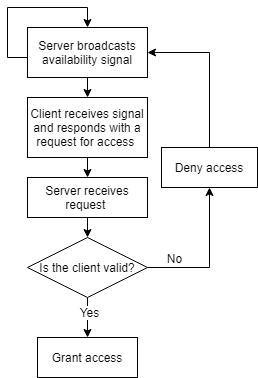}
\caption{Logical flow of server in processing access requests from clients.\label{flow}}
\end{figure}

The server provides the network services while the client is the device that attempts to request access to the network and its data. The server organizes the permissions that govern each client within the network to ensure that each endpoint is safe. Before carrying out any transaction between clients, the server needs to do a security check. This sequence allows complete control over any form of data flow within the network. A client must first have its request for access approved by the server before the data becomes visible. This verification step is through a packet exchange called a handshake. For this architecture, we use a three-point handshake as the main design flaw of the proposed platform. Its flow starts with a device sending its handshake packet to the server requesting to connect. The server then takes this packet and attempts to verify it. Upon its verification, the network processes the message. However, if it cannot be determined, then no access is given. A logical flow chart presenting how the server processes access requests from clients is in Fig.~\ref{flow}.

\subsubsection{Client}
The client is the device that attempts to request access to the network data by sending the handshake packet to the server. Any device that can request access to the provided services from a server is considered a client. Most of the time, the client waits for the server response for every request it sends. However, before that, it must first detect which servers are available through an initial handshake packet. The process continues with the client sending an access request to the source of the initial message. Then, the client waits for a response from the server. If a response from the server is received, the client checks if the network granted its request. If so, then the client is now free to use the services of the server.

\subsection{Implementation}
\begin{figure}[t!]
\centering
\includegraphics[width=\columnwidth]{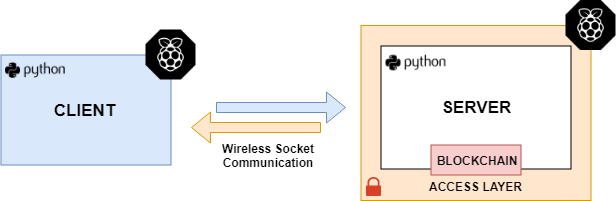}
\caption{Visual abstraction of proposed design setup for the communication between client and server.}
\label{setup}
\end{figure}

This section highlights how we implemented each part and the processes that carry out the previously highlighted logic. A visual representation of the setup described in this section is in Fig.~\ref{setup}. We use Raspberry Pis for both server and client. This setup simulates the interaction between the two. Also, they are connected wirelessly through a wireless network. In terms of data transmission, the Raspberry Pis will communicate through a socket module via serial communication. This method of sharing information will be in charge of the packet exchange between the server and client. As for Operating Systems (OS) and other software-related configurations, the Pis will be running on a Raspbian. Also, we programmed all the scripts in Python. The rest of the information on our implementation is composed of subsections. It highlights the structures and setups of the platform. 

\subsubsection{Cryptographical Setup}
Initially, the idea was to create secure communication between the server and client. The server was programmed to check for a unique serial ID that only the device and the server knows. This design grants the server the ability to filter devices based on the authenticity of this ID. We took the serial ID from the serial number of each unit, which is unique for each hardware. Thus, making the serial key a potential passkey for the device. However, the main issue is that anyone can find the serial number by just looking at the hardware specifications of each device. At the same time, anyone that can access the server can get a list of all the serial numbers. They can then fake a handshake with any serial number to gain access to any of the endpoints. Therefore, we used public-private key cryptography paired with one-way encryption to ensure authenticity. 

The script used a Linux library called Crypto which contained different ciphers to encrypt the keys and the messages sent by both the server and the client. Also, we incorporated a combination of Advanced Encryption Standard (AES), Rivest–Shamir–Adleman (RSA) ciphers and, a one-way hashing in the implementation. AES and RSA were used to encrypt parts of the message. We used one-way hashing to generate irreversible encryptions to allow the script to check if there is tampering in the packets. We also used RSA ciphers to create the public and private key pairs. These keys encrypt messages between the server and the device, which secures a route for data to travel. We embedded digital signatures of each device in each packet to serve as a watermark. This design allows the server to verify the data in terms of the structure of its signature. The cryptographic tools are; the serial ID, public key, private key, and the cryptographical tools mentioned. In this design, to make sure this serial ID is safe, one-way hashing is used.

\subsubsection{Packet Structure}
\begin{figure}[t!]  
\centering
\includegraphics[width=0.9\columnwidth]{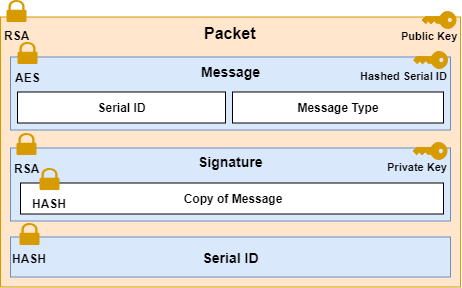}
\caption{Structure of packets that are exchanged between server and client.\label{packet}}
\end{figure}

The packet between the server and the client follows a unique structure to allow secure exchanges. Our proposed platform includes a packet structure designed to maximize the cryptographic capabilities towards securing the data shared between two devices. We programmed each one to contain an encrypted version of the message, a signature, and the serial ID of the device. A visual representation of the different encrypted layers composing this proposed structure, is shown in Fig.~\ref{packet}. As mentioned in the previous section, we used a combination of RSA, AES, and one-way hashing to create a secure architecture. The packets sent by the client and server are protected by; a serial ID, the public key that each entity knows, and the private key that is only known to the device that holds it. The following paragraphs provide a detailed description of the encryption process of each packet.

Firstly, the message is encrypted using the serial ID of the device that is one-way hashed. This design protects the data from being read by anyone who does not know the serial ID and is unaware of the hashing algorithm used to generate the encryption key. Within this message is the type of the message along with the device's serial ID. Since this implementation only focuses on the access level of the exchange, this is the only information in the packet. 

Next, the sending node generates the signature through the RSA cipher by using the private key of the packet's source and a hashed copy of the original message. This receiving node then verifies the signature by confirming the integrity of the received message with the hashed copy store with the signature. In addition, the receiver can only open the signature with the public key of the sender. We programmed this signature along with the message and the sender's serial ID. 

\begin{figure}[t!]
\centering
\includegraphics[width=0.4\columnwidth]{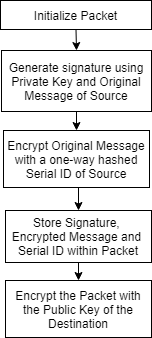}
\caption{Logical flow of packet creation in client and server.}
\label{create}
\end{figure}

The sender encrypts the public key of the destination of its packet before it is transmitted. This process makes it so that the message is only readable with the private counterpart of the public key that encrypted it. A visual representation of this encryption process is through a logic flow diagram in Fig.~\ref{create}.

\subsubsection{Server and Client}
The server is a python class that contains the encryption and decryption methods used to process, send and receive messages. Within the server is a programmed permissioned blockchain implementation containing all the devices that can access the server. In terms of initialization, a script written in Python will build the blockchain. We constructed it as a linked list of blocks that are connected cryptographically. Within each block will be the serial IDs and keys of devices either blocked or allowed in the IoT network. We stored these values transparently through the blockchain since both are visible without compromising the security of the messages. This design is because only with a private key can this information be used to decrypt any message. As for accessing the blockchain, the server will search through its blocks for the serial ID and public key pair that it obtains from the packet. The server can only read the contents of the message if the source is in the ledger. Otherwise, the message is blocked. Also, the network denies any further communication with that device.

The client is a python class that the script initializes by providing a public key and serial ID. Then, the device can be paired by the network using the public key and serial ID. Just like the server, it contains the same encryption and decryption methods to process the messages. The server and the client scripts can be initialized and ran using a socket method. This connection establishes a serial communication between the two. This bridge is created by assigning a static Internet Protocol (IP) address to the server and having the devices call this IP using the socket library provided in Python.

\subsubsection{Communication Setup}
The implementation is initialized by pairing a device with a server. The previous section mentions that the flow of the interaction was a three-point handshake. Due to the constraints, we placed the public key of the server in the client for the initial handshake. Also, we stored the public key of the client and its serial ID within the blockchain. As a result, the server treats the blockchain as a ledger. The client then sends an access packet to the server following the packet structure. The server receives this message via socket programming and attempts to decrypt it. 

The packet is decrypted systematically by doing its encryption process in reverse. First, the server decrypts the packet using its private key. This design is due to RSA encryption which protects any message from being opened unless the private counterpart of the public key that encrypted the packet is available to the server. The server then takes the serial ID within the decrypted packet and searches the blockchain for any registered device. If there is a match, the server can then verify the source. 

\begin{figure}[t!]
\centering
\includegraphics[width=0.65\columnwidth]{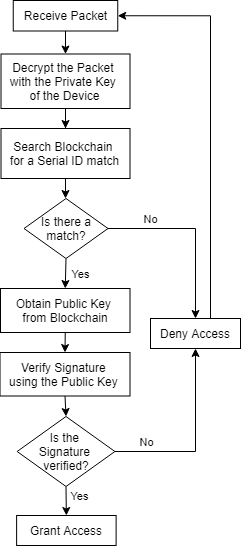}
\caption{Logical flow of packet decrypting and processing done by server.}
\label{decrypt}
\end{figure}

With the public key of the source, the recipient can verify the signature. It will be allowed if the message is accurate. Otherwise, it denies it if the message has been tampered with before it reached its destination. Once the signature is verified, the message is now considered valid. For this implementation, it can only check if the packet is an access type. If so and its source is proven authorized, the server sends back a message using the same packet structuring. It notifies the device that it has access. A visual representation of this decryption process is through a logical flow diagram in Fig.~\ref{decrypt}.

\begin{figure}[t!]
\centering
\includegraphics[width=\columnwidth]{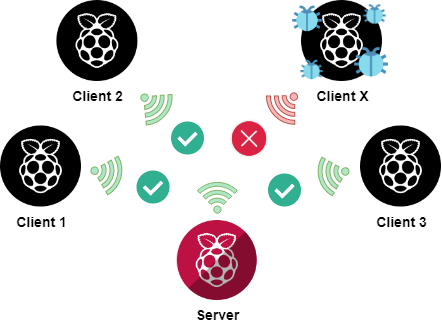}
\caption{Visual abstraction of the simulated malicious attack to the server.\label{attack}}
\end{figure}

With this design, attacks can also be simulated by having multiple clients attempt to connect while not being registered to the server. A visual abstraction that models a simulated malicious attack by the proposed platform is in Fig.~\ref{attack}. We tested this implementation with devices that are either included in the blockchain or not.

\section{Security Analysis and Evaluation} \label{eval}
Improving the security of Fog-IoT networks is the primary purpose of this implementation. Each technique and design choice that this paper highlight makes up a proposed solution that focuses on the data transactions between a client and the server of a Fog-IoT network. We carried out the implemented solution in a controlled environment. The simulation consists of having the user attempt to access data from the server by sending handshake packets. The server then responds accordingly based on the digital signatures structured within the message. Within this design, we put constraints in place. This design choice is due to the limitations of the hardware and software used. However, even with these constraints, the design should still be made sure to meet an adequate level of security. 

There are various types of modelling methodologies that evaluate and analyze the design. Some examples of these are; System Theoretic Process Analysis-Security (STPA-Sec), Hazard and Operability (HAZOP), and Operationally Critical Threat, Asset, and Vulnerability Evaluation (OCTAVE). Lastly, there is STRIDE, which is an acronym for Spoofing, Tampering, Repudiation, Information Disclosure, Denial of Service (DoS), and Elevation of privilege~\cite{model-method}. Each method is selected based on the aspect of the system that it specializes in evaluating and modelling. In~\cite{stpa-sec}, they make use of the STPA-sec model for a case study of their proposed air refuelling system. This modelling methodology focuses on analyzing the safety of the system. In~\cite{hazop}, they use the HAZOP model to evaluate the reliability of their proposed Heat Treatment System. This methodology focuses on the hazard and operability of the system. In~\cite{sardjono}, they use the OCTAVE method to evaluate the information system security of a Deutsche bank. This method focuses more on operationally critical threats and assets related to the system. In~\cite{al-asif}, they use the STRIDE model to evaluate the security in a telesurgery system. This model specializes in using different types of cyber threats in determining the level of vulnerability of a design. 

Although each method has its strengths in terms of covering different aspects of the system, the STRIDE threat model presents the best means of describing vulnerabilities and identifying design exploits~\cite{fuzzy-stride}. Meanwhile, the other three methods focus more on the hazards and risk analysis of the system. Compared to other methods, STRIDE provides a simpler model for evaluation that requires lesser complex tests. Therefore, to fully evaluate the strengths and weaknesses of our design, we chose to use the STRIDE model~\cite{stride}. This threat model is an evaluation tool that classifies security threats into six general categories. It models potential vulnerabilities within a system. The evaluation then uses this information to gauge the ability of the current design to defend itself from attacks categorized under each threat~\cite{stride-model}.  

The following enumerates the results of the analysis of our proposed solution using the STRIDE model:

\begin{enumerate}
    \item Spoofing - It is the process of impersonating a known user in a network. With the reinforced encryption of the keys used by each user within our proposed platform, successfully impersonating one will be a challenge. Due to the number of keys required for a device to establish a connection, the process of obtaining a user's key is complex. 
    
  A device must first know the public key and serial ID of the server to establish a connection. The device must then have a pre-registered public-key and serial ID pair. Finally, the device must have the private counterpart of the registered public key. 
    
    Since the private and public keys can only be created by the network simultaneously, a private key remains secure. Unless the owner of the key discloses this key or an attacker manages to gain access to the device. This condition creates a complex system that even though spoofing is still possible with attacks of varying intensities, penetrating the server will still be a challenge. By using permissioned blockchains to store these keys, it secures the data against targeted attacks. As a result, this combination of technologies improves the security of the IoT network. 
    
    In terms of other variables, there is the number of devices connecting to the network. It should not impact the consistency of this design. Since the public-key and serial ID pair is already registered, the platform can filter through all the messages that it receives. Although it might slow down the network throughput, it would still be able to sift through each packet to catch any spoofing attacks. Therefore, in terms of the spoofing evaluation, the proposed design had countermeasures against it. However, it does not guarantee complete security against its threats. 
    
    \item Tampering  - It is an act of altering the contents of a message or a system without authorization. Anyone who can access the Raspberry Pis can change the scripts of the client and server. Therefore, since the code is open for editing, users can directly change the device's identity and reinitialize the code. 
    
    However, once a message is received, no one can tamper with its contents. With the digital signature, the system can check the message's integrity with a hashed version of the original message as the device places it into the packet. By doing so, the server can detect tampering. Also, it can discard flagged messages. This feature creates a filter that acts against tampering. It can use the access layer and decide which packets to discard. In addition, an attacker cannot tamper with the blockchain if the network has already deployed the server. The immutable nature of the blockchain defends the server against any tampering attacks. Most changes to the implementation can be during the initialization of the devices. It will be harder to tamper with the parameters used in the initialization after running scripts that drove the different aspects of the implementation. 
    
    Similar to the spoofing evaluation, increasing the number of devices should not impact the ability of the blockchain to resist tampering. Usually, it is the connection that bottlenecks the users that access the server. This occurrence might cause the processing of data to slow down. However, each user still receives equal security against tampering. As we use the tamper-proof and immutability blockchains, the proposed platform can act against tampering attacks. 
    
    \item Repudiation - Repudiability is the state where it is possible to verify a transaction. A message is repudiated once its receiver authenticates it. In terms of the implementation, repudiation is made possible with the digital signatures in each packet. Digital signatures contain a hashed copy of the original message before the sender transmits it. Once the message is received, the recipient can cross-check its contents with is in the signature. Then, the server investigates it for any inconsistencies with the signature. 
    
    Since the copy of the original message is protected by hashing, tampering with it is close to impossible. Also, if an attacker manages to alter the message, the signature is there to detect it. Also, the network can only generate signatures with a private key. Therefore, it will be hard for an attacker to replace the signature in a packet. Multiple devices that send messages will not impact the server's ability to check the authenticity of a transaction. Since all received packets still go through the server, it will verify each message without exceptions. 
    
    Therefore, with a digital signature tied to each device, messages can be verified by the network. This functionality adds to the preservation of the repudiability of the data transactions within the Fog-IoT network. Also, by using the automatability of the blockchain, we can further improve the ability of the network to detect data tampering.
    
    \item Information Disclosure - There are three crucial pieces of information stored by each device. These are the private key, public key and serial ID. The nature of the design requires the sharing of the public key and serial ID with the server. This feature means the users who have access to the server can see two of the three pieces of information. However, the private key is never disclosed to any user apart from its generator. This key secures the privacy of even the messages between the server and the client. The private key of the recipient protects the packet. If it is secure, then the content of the message is also safe. As a result, breaches in privacy in this design become highly unlikely. Therefore, by withholding an essential key to hiding data from attackers, private information can remain secure. 
    
    \item Denial of Service - Denial of service (DoS) focuses on the ability of the server to do its purpose. This implementation focuses on a secured access layer and the communication between the server and the client. Thus, the nature of blockchains can address this threat. Due to their decentralized structure, we can implement a distributive network using Fog-IoT.
    
    With a distributed consensus, the design can have the ability to adjust against massive amounts of incoming access requests. As a result, the server gets proper protection against DoS attacks using the permissioned blockchain. Due to the decentralized and distributive nature of Fog-IoT networks, increasing the number of devices will be easier to manage through resource reallocation. Therefore, the server can defend itself from DoS attacks that scale based on the number of attackers. However, this is not the same with the clients. Without a proper filtering protocol within a client's device, more complex DoS attacks that target the users can be an issue for our design. Therefore, we can focus on creating a rate-limiting protocol within the server for future iterations. Also, it can be a network that detects DoS attacks that target its users.
    
    \item Evaluation of Privilege - In the current iteration of this design, there are only two levels; has access and has no access. A device cannot elevate their privilege unless they are in the ledger of the server. Just as the previous aspect, since this implementation only focuses on the authentication layer, defensive measures against multiple layers of authority within the network are missing. As a result, our proposed design only provides a basic allotment of security towards the client-side of this aspect. On the other hand, servers provide no further defences against devices that can access their data. As a result, data restricted to administrators can be vulnerable to all trusted users. This development might also be a growing concern as the network adds more devices to the server. The result will be more server data will be exposed to more users. However, the improvement of the server's architecture can address this issue. Due to blockchain programmability and automatability, we can establish better access levels with permissioned blockchains. The result is better security rankings and defences. Therefore, further improvements to the blockchain structure can create a better authorization protocol for the network.

\end{enumerate}

A summary of the evaluation of the design in terms of the STRIDE model is in Table~\ref{stride-table}. The proposed platform shows its ability to defend itself against threats such as spoofing, tampering, repudiation, and information disclosure. However, there are vulnerabilities found against DoS and evaluation of privilege attacks. Based on the analysis, we found clients to be more prone to DoS attacks. Meanwhile, the current setup of the server is not secure against privilege attacks. Also included within the evaluation is further discussion on how to improve the current iteration. Furthermore, variables such as; the number of devices and intensity of attacks were in the enumeration of each threat. Overall, the design shows promise for being able to cover most of the categories while having the potential of being able to solve the rest. 

\begin{table}[t!]
\fontsize{10}{12}\selectfont
\centering
{
\begin{tabular}{|p{1cm}|p{0.5cm}|p{0.5cm}|p{0.5cm}|p{0.5cm}|p{0.5cm}|p{0.5cm}|}\hline
Aspect  & S & T & R & I & D & E\\\hline
Client & X & X & X & X & - & X \\
Server & X & X & X & X & X & - \\
\hline
\end{tabular}
}
\caption{STRIDE model evaluation of the proposed design.}
\label{stride-table}
\end{table}

\section{Conclusions} \label{con}
In this work, we proposed a platform that aims to strengthen security within a Fog-IoT network. With the introduction of endpoints through the local servers, the network is left vulnerable to malicious attacks. As a result, data within the server is at risk. We proposed a platform to protect the data transactions between client and server from malicious attacks. Our design introduces a trusted authentication system using permissioned blockchains. This technology creates a systematic means of filtering devices that connect to the IoT network. Also, we plan to reinforce the security of the data transmitted within the platform by using a combination of public keys and one-way hashing. These cryptographic techniques create a data packet structure to protect the data shared between the server and its clients. By combining these technologies and techniques, we have a Fog-IoT network platform for system administration.  

To test the feasibility of the security of our proposed platform, we cross-examined it against a STRIDE threat model. The results showed that the packets transmitted between the client and server using the proposed platform passed the spoofing, tampering, repudiation, and information disclosure aspects. With the private blockchain and the packet structure, we created a secure design for Fog-IoT networks. However, the model also showed that the platform could not address DoS attacks against the client and the elevation of privilege in servers. As a result, we proposed areas of the design that we could improve for future iterations. Although we could not cover all aspects of the threat model, our platform showed its ability to handle a diverse collection of security threats. By integrating permissioned blockchains with a public key encrypted packet structure, we achieve a great degree of coverage in Fog-IoT security. Therefore, we can see the overall feasibility and potential of the platform in securing the system administration of a Fog-IoT network.

\bibliographystyle{IEEE}
\bibliography{IEEEabrv,secbib}
\end{document}